\documentclass{aa}
\usepackage{txfonts}
\usepackage{graphicx}

\usepackage{color}

\begin{document}

\title{The kinematics of  Small Magellanic Cloud star clusters}

\author{Andr\'es E. Piatti\inst{2,3}\thanks{\email{andres.piatti@unc.edu.ar}}}

\institute{Instituto Interdisciplinario de Ciencias B\'asicas (ICB), CONICET-UNCUYO, Padre J. Contreras 1300, M5502JMA, Mendoza, Argentina;
\and Consejo Nacional de Investigaciones Cient\'{\i}ficas y T\'ecnicas (CONICET), Godoy Cruz 2290, C1425FQB,  Buenos Aires, Argentina\\
}

\date{Received / Accepted}

\abstract{
We report results of proper motions of 25 known Small Magellanic Cloud (SMC) clusters (ages $\sim$
1 - 10 Gyr old)  derived from {\it Gaia} EDR3 data sets. When these mean proper motions are 
gathered with existent radial velocity measurements to compose the clusters’ velocity vectors, we found 
the parameter values of a rotation disk that best resemble their observed motions, namely:
central coordinates and distance, inclination and position angle of the line-of-node, proper motion in right ascension and declination and systemic velocity, rotation velocity and velocity dispersion. The
SMC cluster rotation disk seems to be at some level kinematically synchronized with the
rotation of field red giants recently modeled using DR2 data sets. Such a rotation disk is seen in the 
sky as a
tilted edge-on disk, with a velocity dispersion perpendicular to it twice as big as that in the plane of
the disk. Because the direction perpendicular to the disk is nearly aligned with the Magellanic Bridge,
we interpret the larger velocity dispersion as a consequence of the SMC velocity stretching
caused by the tidal interaction with the Large Magellanic Cloud. Rotation alone would not seem
sufficient to explain the observed kinematic behaviors in the SMC.} 
 

\titlerunning{SMC star cluster kinematics}

\authorrunning{Andr\'es E. Piatti}

\maketitle

\markboth{Andr\'es E. Piatti: }{SMC star cluster kinematics}

\section{Introduction}

During the last years several efforts have been made in order to accurately trace
the internal kinematic behavior of the Small Magellanic Clouds (SMC) 
\citep[see, e.g.,][and references therein]{niederhoferetal2018,zivicketal2018,diteodoroetal2019}.
Radial velocities-only analysis, surveys based on proper motion measurements and some
other studies that complemented both kinematics information arrive at some extent to different
results. Although there is a general consensus about the presence of tidal effects by the
Large Magellanic Cloud (LMC), the magnitude of such a tidal disruption turns out to be
different, depending on the stellar population examined \citep{deleoetal2020,zivicketal2020,niederhoferetal2021}. Furthermore, an ordered SMC 
disk rotation kinematics is nowadays under debate.  Indeed, while \citet{deleoetal2020} 
and \citet{niederhoferetal2021} do not detect any signature of rotation, \citet{zivicketal2020} find 
a moderate rotational motion.

As far as we are aware, star clusters have not been used as tracers of the SMC internal
kinematics. However, they are advantageous from different points of view. Their
mean radial velocities and proper motions come from the average of measurements of
several cluster members, resulting in values based on a robust statistics. They are
also suitable representatives of the internal motion of the SMC, as compared with the 
mean radial velocities and proper motions of stars aligned along different line-of-sight across the SMC. 
This is also valid even in the case of dealing with different field stellar populations analyzed 
separately. On the other hand, star clusters are accurately aged, so that the internal motion of the
SMC can be easily linked to the star cluster metallicities, thus providing a comprehensive
framework for our understanding of the formation and chemico-dynamical evolution history of 
the galaxy.

In this work, we uncovered for the first time the ongoing internal motion of the SMC as seen
by its star cluster population older than $\sim$ 1 Gyr. We built a 3D picture of their movements
from the {\it Gaia} Early Data Release \citep[{\it Gaia} EDR3][]{gaiaetal2016,gaiaetal2020b} and
radial velocities gathered from the literature to construct their space velocity vectors. 
By analyzing these space velocity components we find that the SMC can be described
as a rotating edge-on disk, which is experiencing a stretching process nearly perpendicular 
to it, in direction toward the LMC. The Letter is organized as follows: in Section 2, we describe 
the collected {\it Gaia} EDR3 data sets for the star cluster sample. In Section 3, we perform a 
rigorous selection of cluster members and estimate the mean star cluster proper motions.
Section 4 deals with the analysis of the space velocity components and discusses the
results to the light of recent  finding in this field.

\section{Collected data and mean cluster proper motions}

We started by searching the literature for  radial velocities studies of SMC star clusters, and
 found that  \citet{dh98}, \citet{parisietal2009}, \citet{parisietal2015}, and  \citet{kamannetal2018}
derived mean radial velocities for a total of 31 clusters. 
Then, we used the cluster catalog compiled by \citet{bicaetal2020} to extract right ascension, declination,
size, age and metallicity for those star clusters with radial velocity studies  \citep[see, also,][]{piatti2021}. 
From {\it Gaia} EDR3\footnote{https://archives.esac.esa.int/gaia.}, we extracted
parallaxes ($\varpi$), proper motions in right ascension (pmra) and declination (pmdec),
excess noise (\texttt{epsi}) and significance of excess of noise (\texttt{sepsi}), and $G$,
$BP$ and $RP$ magnitudes for stars located within a radius equal to three times the cluster radius 
from the respective cluster center. We limited our sample to stars with proper motion errors $\le$ 0.1 
mas/yr, following the suggestion by \citet[][and reference therein]{piattietal2019}, who found 
that the larger  the individual proper motion errors of LMC stars, the larger the derived errors of the 
mean LMC
cluster proper motions. We minimized the presence of foreground stars by favoring distant stars,
i.e., $|\varpi|$/$\sigma(\varpi)$ $<$ 3, and pruned the data with
\texttt{sepsi} $<$ 2 and \texttt{epsi} $<$ 1, which define a good balance between data quality and 
number of retained objects for our sample \citep[see also,][]{ripepietal2018}.
We built  color-magnitude diagrams (CMDs) of the selected stars and found a limiting
magnitude of $G$ $\la$ 18.0 mag. At the mean SMC distance, \citep[see, e.g.,][]{graczyketal2020},
this limit corresponds nearly to the base of the red giant branch.

\begin{figure}
\includegraphics[width=\columnwidth]{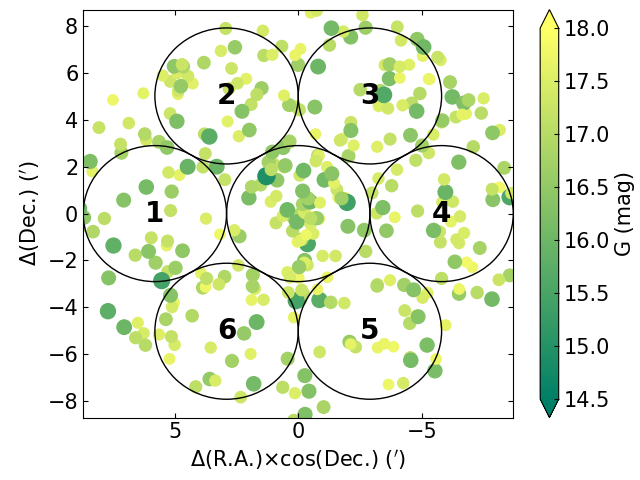}
\caption{Selected {\it Gaia} EDR3 stars distributed in the field of NGC\,339. The 
cluster circle
and 6 adjacent star field circles of equal size are also drawn. The size of the symbols 
is proportional to the star brightness ($G$ mag).}
\label{fig:fig1}
\end{figure}

In order to statistically clean the cluster vector-point diagrams (VPDs) from the
SMC field star contamination, we employed a procedure originally devised to decontaminate 
cluster CMDs \citep{pb12},  which was satisfactorily
applied in cleaning CMDs of star clusters projected toward
crowded star fields  \citep[e.g.,][and references therein]{p17a,p17b,p17c} and affected by differential 
reddening \citep[e.g.,][and references therein]{p2018,petal2018}. The method basically 
consists in using a circular star field located close to the cluster field -also delimited by a circle-, 
then building its respective VPD, and finally subtracting it from the cluster VPD.
Fig.~\ref{fig:fig1} illustrates, as an example, the distribution of   six star fields
around NGC\,339, superimposed on the stars that comply with the above selection criteria.
The cluster circle, centered on the cluster and with a radius equal to the cluster
radius, and the  six adjacent star field circles are of the same area.  The six
adjacent star fields are used one at a time, thus repeating the cleaning procedure
to improve the statistics.

We subtracted from  the cluster field a number of stars equal to that in the adjacent 
star field.  
The method  defines boxes  in the star field VPD
centered  on each star; then  superimposes all the boxes
on the cluster VPD, and chooses one star per box to subtract.
With the aim of avoiding stochastic effects
caused by very few boxes distributed in  some VPD regions, 
variable box sizes were used. In the case that more than 
one star is located inside  a box, the closest one to its center is 
subtracted. 
The positions in the cluster field of the subtracted stars were chosen
randomly within small boxes of 10$\%$ the cluster radius per side. 
We iterated this search up to a thousand times throughout the cluster field.
The proper motion errors  of the stars in the cluster field were also taken into account 
while searching for a star to subtract from the cluster VPD. With that purpose, we 
iterated up to a thousand times  the search for each  VPD box, allowing the stars in the
cluster VPD to vary their proper motions within an interval of $\pm$1$\sigma$,
where $\sigma$ represents the errors in their proper motions. For the sake of the
reader, Fig.~\ref{fig:2} shows the resulting cleaned cluster
VPD and cluster field, respectively, using different adjacent star fields.

We assigned a membership probability to each star  in the cluster field
as $P$ ($\%$) = 100$\times$$S$/6, where $S$ represents the number of times the
star was not subtracted after the six different VPD cleaning executions.
 With that
information on hand, we built Fig.~\ref{fig:fig3} to illustrate the VPD of all the selected
stars located in the field of NGC\,339. Stars with
different $P$ values were plotted with different colors. We checked the
distribution of the stars along the cluster red giant branch using the {\it Gaia} EDR3
photometry and kept only those with $P>$ 50$\%$. Table~\ref{tab:tab1}
lists 25 SMC clusters that have passed all the cleaning and selection procedures,
and the respective number (N) of stars. 
Other clusters (HW\,40, 47, 67, L\,11, 13, 108) were also analyzed and no star
remained at the end of the above analysis.

We finally performed a maximum likelihood statistics \citep{pm1993,walker2006} in 
order to estimate the mean  proper motions and dispersion for the cluster
sample. We optimized the probability $\mathcal{L}$ that a
given ensemble of stars with 
proper motions pm$_i$ and errors $\sigma_i$ are drawn from a population with mean
proper motion $<$pm$>$ and dispersion  W, as follows:\\

$\small
\mathcal{L}\,=\,\prod_{i=1}^N\,\left( \, 2\pi\,(\sigma_i^2 + W^2 \, ) 
\right)^{-\frac{1}{2}}\,\exp \left(-\frac{(pm_i \,- <pm>)^2}{2(\sigma_i^2 + W^2)}
\right) \\$

\noindent where the errors on the mean and dispersion were computed from the respective covariance matrices. We applied the above procedure for pmra and 
pmdec, separately. The resulting mean values are listed in Table~\ref{tab:tab1}.

\begin{figure*}
\includegraphics[width=\columnwidth]{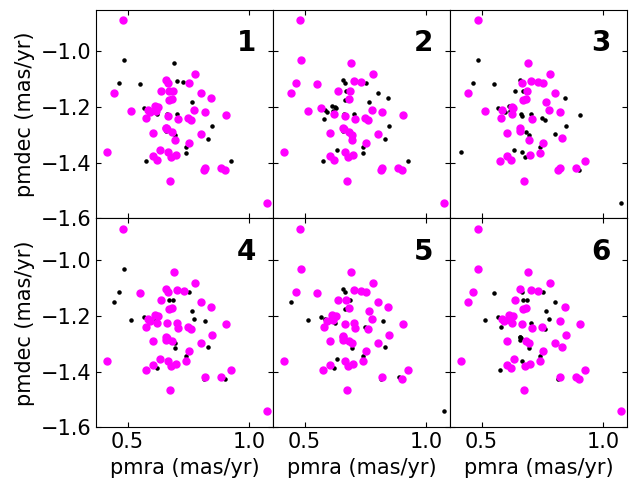}
\includegraphics[width=\columnwidth]{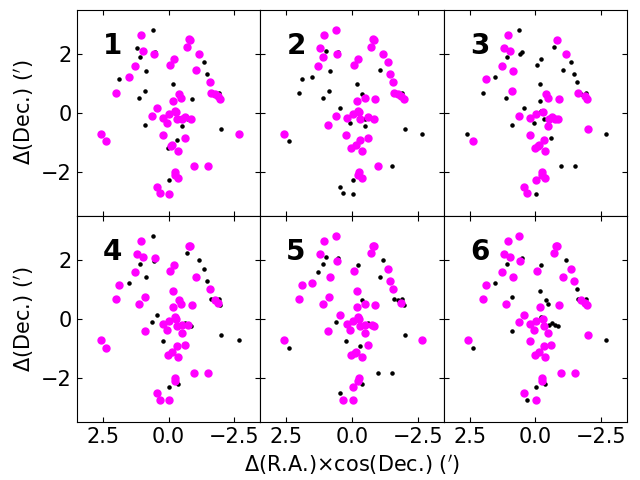}
\caption{VPD (left panels) and cluster field (right panels) of NGC\,339. The labels
indicate the adjacent star field used to clean the cluster VPD. Magenta symbols
represent the stars that remained unsubtracted after the cleaning procedure.}
\label{fig:fig2}
\end{figure*}

\begin{figure}
\includegraphics[width=\columnwidth]{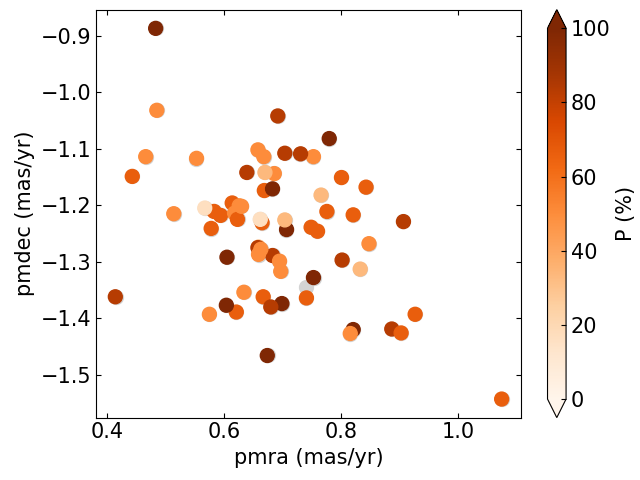}
\caption{VPD for selected {\it Gaia} EDR3 stars distributed in the field of 
NGC\,339, colored according to the assigned membership probability.}
\label{fig:fig3}
\end{figure}

\begin{table*}
\caption{Proper motions of SMC star clusters.}
\label{tab:tab1}
\begin{small}
\begin{tabular}{lccclccclccc}\hline\hline
Cluster & pmra & pmdec & N & Cluster & pmra & pmdec & N & Cluster & pmra & pmdec & N \\
          & (mas/yr) & (mas/yr) &  &  & (mas/yr) & (mas/yr) & &  (mas/yr) & (mas/yr) & \\\hline
BS\,121&	0.654$\pm$0.081  & -1.143$\pm$0.046  &   8 & L\,9	 &0.419$\pm$0.086  & -1.159$\pm$0.086  &   3& L\,112  &1.133$\pm$0.058 &	-0.971$\pm$0.043  &    3\\
HW\,84 &	1.210$\pm$0.034   &-1.201$\pm$0.059   &  2 & L\,12&  0.454$\pm$0.076 &  -1.204$\pm$0.072  &   3 & L\,113  &	1.287$\pm$0.033  &  -1.221$\pm$0.022 & 17\\
HW\,86 &	1.191$\pm$0.112   &-1.276$\pm$0.154   &  2 & L\,17& 0.531$\pm$0.060 &	-1.215$\pm$0.054   &    9& NGC\,121 &0.344$\pm$0.025 &	-1.196$\pm$0.022  & 33\\
L\,1	  &	0.575$\pm$0.011   &-1.520$\pm$0.014  & 39 & L\,19 &0.634$\pm$0.112   & -1.331$\pm$0.062 &	  3& NGC\,339	 &	0.684$\pm$0.019 &	-1.256$\pm$0.018  &  37\\
L\,4	 &	0.437$\pm$0.044 &	-1.290$\pm$0.057   &   5 & L\,27 & 0.724$\pm$0.043 &	-1.427$\pm$0.066   &   11& NGC\,361&	0.796$\pm$0.039 &	-1.221$\pm$0.035  & 17\\
L\,5	 &	0.529$\pm$0.040  & -1.345$\pm$0.081   &  4 & L\,58 & 0.472$\pm$0.083  & -1.322$\pm$0.075 &	 1& NGC\,419	  &	0.783$\pm$0.063 &	-1.230$\pm$0.029  &   17\\
L\,6	 &	0.510$\pm$0.060 &  -1.280$\pm$0.042  &    7 & L\,68 & 0.711$\pm$0.038  & -1.225$\pm$0.031  &   15& NGC\,643	 &	1.259$\pm$0.099   &  -1.301$\pm$0.034 &    3\\
L\,7	 &	0.445$\pm$0.031  & -1.273$\pm$0.079   &  4 & L\,106 &1.125$\pm$0.078  & -1.313$\pm$0.041  &   4  &  & & & \\
L\,8	 &	0.545$\pm$0.023   &-1.287$\pm$0.024 &	39 & L\,110 & 0.816$\pm$0.033  & -1.182$\pm$0.020 & 	 8 &  & &  & \\\hline
\end{tabular}
\end{small}
\end{table*}

\section{Analysis and discussion}

\citet{zivicketal2020} obtained the best-fit parameters of a rotation disk 
which represents the kinematics of SMC field red giants by using {\it Gaia DR2} data,
radial velocities from the literature and the transformation equations (9), (13) and (21) 
in \citet{vdmareletal2002} \citep[see, also,][]{piattietal2019}
as a function of the position angle, measured from north to east. We followed here 
a similar procedure by searching for the kinematic model that best represents the
motion of our cluster sample. The modeled rotation disk is characterized
by the right ascension and declination of its center, its distance, radial
velocity and proper motions in right ascension and declination, the inclination
of the disk, the position angle of the line-of-nodes (LON) and the rotational
velocity. Velocity dispersions along the three independent axes were also taken into
account. In order to derive their best representative
values we allowed them to vary in fixed ranges, as shown in Table~\ref{tab:tab2}.
For each possible combination of the parameter values,
we computed the rotational velocity components $v_1$, $v_2$ and $v_3$
\citep[see][]{vdmareletal2002} as a function of the position angle
of a point placed on a
disk at a distance $s$ to its center. From the grid chosen in the parameter space, we
generated more than 5.3$\times$10$^8$
different curves for $v_1$, $v_2$ and $v_3$, respectively, from which we
extracted the values corresponding to the clusters' position angles.

\begin{table*}
\caption{SMC disk rotation parameters.}
\label{tab:tab2}
\begin{tabular}{lccc}
\hline\hline
Parameter & Parameter range & Parameter step & Best representative values \\\hline
SMC center right ascension ($\degr$): 		& 12.50 - 13.35 & 0.05  & 13.30$\pm$0.10 \\
SMC center declination ($\degr$):		& -73.20 - -72.70 & 0.05 & -72.85$\pm$0.10 \\
SMC center distance (kpc):				& 58.0 - 64.0 &1.0 & 59.0$\pm$1.5\\
SMC center pmra (mas/yr):				&0.6 - 0.9 & 0.05 & 0.75$\pm$0.10\\
SMC center pmdec (mas/yr):			&-1.30 - -1.15 &0.02 & -1.26$\pm$0.05\\	
SMC center systemic velocity (km/s):		&130.0 -160.0 &5.0 & 150.0$\pm$2.0\\
SMC disk inclination ($\degr$):			&0.0 - 90.0 & 5.0 & 70.0$\pm$10.0 \\
SMC disk position angle LON ($\degr$):	&0.0 - 360.0 &20.0 & 200.0$\pm$30.0\\
SMC disk rotation velocity (km/s):		&0.0 - 30.0 & 5.0 & 25.0$\pm$5.0 \\\hline
\end{tabular}
\end{table*}

On the other hand, we calculated the 
rotational velocity components
for the cluster sample using the mean cluster radial velocities and proper
motions, and the above combinations of disk parameters  \citep[see][]{piattietal2019}. 
As a matter of units, we
used [km/s] = 4.7403885*$D_o$ [mas/yr], where $D_o$ is the distance to the SMC 
center. We then compared the 25 resulting ($v_1$, $v_2$, $v_3$) vectors with
those calculated above (those corresponding to the disk) by adding
the three velocity component differences ($\Delta$$v_i$, $i$=1,2,3) per cluster (absolute values)
and by deriving the respective average value ($\Delta$$v$) and dispersion 
($\sigma$($\Delta$$v$)) for the whole cluster
sample. We note that although $\Delta$$v$ and 
($\Sigma$$_{i=1}^{i=3}$ ($\Delta$$v_i$)$^2$)$^{1/2}$ have different statistical properties,
the result of using one or another turns out to be very similar.
 We repeated this procedure for each of the above combinations of
parameter values and
looked for the parameter set with the minimum average difference ($\Delta$$v$$_{min}$
$\pm$ $\sigma$($\Delta$$v$$_{min}$)),
which we adopted as the best representation to describe the rotation of the studied
SMC clusters. We then considered  $\Delta$$v$ values in the interval
[$\Delta$$v$$_{min}$-$\sigma$($\Delta$$v$$_{min})$; 
$\Delta$$v$$_{min}$+$\sigma$($\Delta$$v$$_{min})$], and their respective parameter
values. The resulting parameters' ranges were adopted as their uncertainties.
The adopted parameters and their uncertainties are listed in the last column of Table~\ref{tab:tab2}. 

\begin{figure}
\includegraphics[width=\columnwidth]{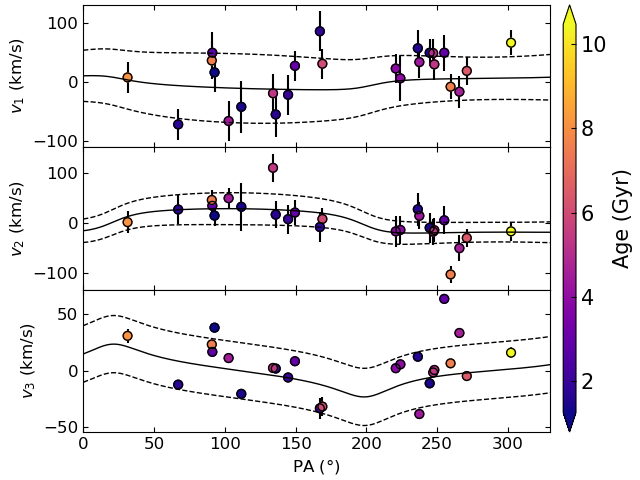}
\caption{Rotation velocity components versus position angle of 25 SMC star clusters. The solid
and dashed lines correspond to the best solution and dispersion for the SMC disk rotation.}
\label{fig:fig4}
\end{figure}

\begin{figure*}
\includegraphics[width=\textwidth]{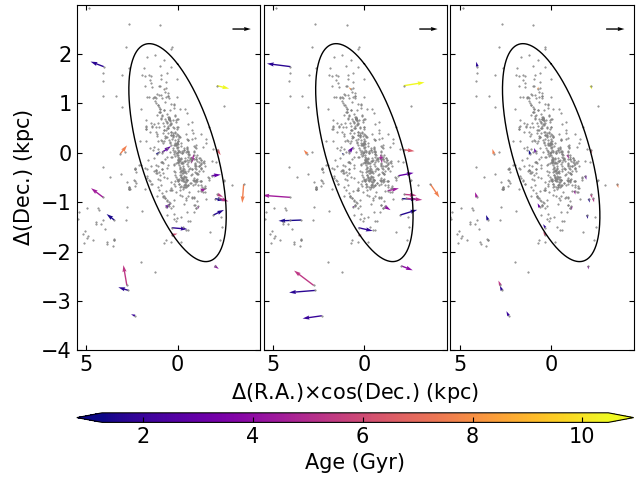}
\caption{Spatial distribution of SMC clusters compiled in \citet{bicaetal2020} (gray dots)
with the elliptical framework adopted by \citet{petal07d} for a semi-major axis
of 3$\degr$ superimposed. The colored age-based vectors represent the residual 
velocity vectors projected on the sky for the studied cluster sample, namely:
the cluster velocity minus the model velocity at its position (left panel), and
the cluster velocity minus the mean SMC motion (middle panel),
and those for the solid lines of Fig.~\ref{fig:fig4} (right panel).}The black arrow 
represents a velocity of 100 km/s.
\label{fig:fig5}
\end{figure*}

Figure~\ref{fig:fig4} depicts the three velocity components as a function of the position
angle for the cluster sample calculated using their mean proper motions and radial velocities and the
best representative rotation disk parameters. We overplotted the 
attained rotation of the SMC disk with a solid line and 
those considering the errors in the inclination of the disk, the position angle of the LON, 
the systemic and transversal velocities of the SMC center and disk velocity dispersion with dashed
lines, respectively. The error bars of the plotted cluster motions account not only for the 
measured errors in proper motion and radial velocity, but also for those from the adopted 
best  solution for the 3D movement of the SMC center, propagated through the
transformation equations and added in quadrature. As can be seen, the cluster motions
relative to the SMC center projected onto the sky resemble that of the rotation of a disk, with some noticeable scatter and some cluster placed beyond the mean rotational pattern.  
\citet{zivicketal2020} point to the need of treating the SMC as a series of different populations 
with distinct kinematics, so that they may provide a better handle on constraining the LMC-SMC interaction, as opposed to measuring averaged systemic SMC properties over multiple stellar populations. We note that 
the cluster ages (1 $<$ age (Gyr) $<$ 10) would not seem to be responsible for the scatter
seen in Fig.~\ref{fig:fig4}, which implies that the use of clusters in the considered age range 
rather than individual stars likely mitigates the concerns raised by \citet{zivicketal2020}.
Instead,  we found that the velocity dispersion plays an important role in dealing with such a scatter
around the mean rotation curves. Particularly, the dashed curves correspond to velocity
dispersions along the right ascension and declination axes and along the line-of-sight 
of 50, 20 and 25 km/s, respectively. This results show that rotation alone is not sufficient 
to explain the observed kinematic behavior in the SMC and, therefore, using a rotating disk 
model for the fitting of the SMC case is subject to uncertainty.
For completeness purposes, we show
in Fig.~\ref{fig:fig5} the velocity vectors projected on the sky.

We compared the resulting parameters with those from some previous works based
on different stellar field populations (red giant branch, RR Lyrae, Cepheids) and
H\,I gas \citep[see Tables 1, 2 and 4 of][ respectively]{zivicketal2020,niederhoferetal2021,deleoetal2020}. We found that the center of the
SMC cluster rotation disk fairly coincides with the so-called optical center
\citep[Right ascension=13.1875, Declination=-72.8286,][]{cetal01}, and is still near to those
determined from SMC stellar components; that from H\,I gas analyses being
clearly different. As far as proper motions of the SMC center is considered, the
present values are in very good agreement with the most recent estimates
 \citep{zivicketal2020,niederhoferetal2021,gaiaetal2020b}, as is also the case of the SMC
systemic velocity \citep{deleoetal2020}. The derived mean distance of the SMC
cluster population agrees with the bulk of previous estimates \citep[][and references therein]{graczyketal2020}, and the representative inclination and position angle of the LON are 
indistinguishable within the errors from those  of the SMC rotation disk 
obtained by \citet{zivicketal2020}.

Such an overall similitude between the resulting SMC cluster rotation
disk (for clusters with ages between 1 and 10 Gyr old) and that of the composite stellar 
field population  suggests that both galaxy components are  at some level
kinematically synchronized. Our findings support the notion of a relatively slow rotation disk 
(rotation velocity of 25 km/s) with respect to the rotation velocity of the LMC 
\citep[$\sim$ 70 km/s,][]{piattietal2019} and the Milky Way 
\citep[$\sim$ 240 km/s,][]{piatti2019}. \citet{zivicketal2020} find a rotation velocity
of 20$\pm$7 km/s for the innermost region with a radius of $\sim$ 2 kpc from the SMC.
Here we confirmed that a similar rotation disk pattern is observed out to $\sim$ 4 kpc 
from the SMC center. In the plane of the sky, the SMC rotation disk can be
roughly represented by a tilted edge-on rotation disk, which is being stretching
in a direction perpendicular to it, nearly parallel to the LMC-SMC connecting
bridge. The stretching signature is witnesses by the dispersion velocity in right
ascension as compared to that along the declination and radial velocity axes
\citep{zivicketal2020,gaiaetal2020b}.

\begin{acknowledgements}
We thank the referee for the thorough reading of the manuscript and
timely suggestions to improve it. 
This work has made use of data from the European Space Agency (ESA) mission
{\it Gaia} (\url{https://www.cosmos.esa.int/gaia}), processed by the {\it Gaia}
Data Processing and Analysis Consortium (DPAC,
\url{https://www.cosmos.esa.int/web/gaia/dpac/consortium}). Funding for the DPAC
has been provided by national institutions, in particular the institutions
participating in the {\it Gaia} Multilateral Agreement.
\end{acknowledgements}


\begin{thebibliography}{29}
\expandafter\ifx\csname natexlab\endcsname\relax\def\natexlab#1{#1}\fi

\bibitem[{{Bica} {et~al.}(2020){Bica}, {Westera}, {Kerber}, {Dias}, {Maia},
  {Santos}, {Barbuy}, \& {Oliveira}}]{bicaetal2020}
{Bica}, E., {Westera}, P., {Kerber}, L. d.~O., {et~al.} 2020, \aj, 159, 82

\bibitem[{{Crowl} {et~al.}(2001){Crowl}, {Sarajedini}, {Piatti}, {Geisler},
  {Bica}, {Clari{\'a}}, \& {Santos}}]{cetal01}
{Crowl}, H.~H., {Sarajedini}, A., {Piatti}, A.~E., {et~al.} 2001, \aj, 122, 220

\bibitem[{{Da Costa} \& {Hatzidimitriou}(1998)}]{dh98}
{Da Costa}, G.~S. \& {Hatzidimitriou}, D. 1998, \aj, 115, 1934

\bibitem[{{De Leo} {et~al.}(2020){De Leo}, {Carrera}, {No{\"e}l}, {Read},
  {Erkal}, \& {Gallart}}]{deleoetal2020}
{De Leo}, M., {Carrera}, R., {No{\"e}l}, N. E.~D., {et~al.} 2020, \mnras, 495,
  98

\bibitem[{{Di Teodoro} {et~al.}(2019){Di Teodoro}, {McClure-Griffiths},
  {Jameson}, {D{\'e}nes}, {Dickey}, {Stanimirovi{\'c}}, {Staveley-Smith},
  {Anderson}, {Bunton}, {Chippendale}, {Lee-Waddell}, {MacLeod}, \&
  {Voronkov}}]{diteodoroetal2019}
{Di Teodoro}, E.~M., {McClure-Griffiths}, N.~M., {Jameson}, K.~E., {et~al.}
  2019, \mnras, 483, 392

\bibitem[{{Gaia Collaboration} {et~al.}(2020){Gaia Collaboration}, {Luri},
  {Chemin}, {Clementini}, {Delgado}, {McMillan}, {Romero-G{\'o}mez},
  {Balbinot}, {Castro-Ginard}, {Mor}, {Ripepi}, {Sarro}, {Cioni}, {Fabricius},
  {Garofalo}, {Helmi}, {Muraveva}, {Brown}, {Vallenari}, {Prusti}, {de},
  {Babusiaux}, {Biermann}, {Creevey}, {Evans}, {Eyer}, {Hutton}, {Jansen},
  {Jordi}, {Klioner}, {Lammers}, {Lindegren}, {Mignard}, {Panem}, {Pourbaix},
  {Randich}, {Sartoretti}, {Soubiran}, {Walton}, {Arenou}, {Bailer-Jones},
  {Bastian}, {Cropper}, {Drimmel}, {Katz}, {Lattanzi}, {van Leeuwen}, {Bakker},
  {Casta{\~n}eda}, {De}, {Ducourant}, {Fouesneau}, {Fr{\'e}mat}, {Guerra},
  {Guerrier}, {Guiraud}, {Jean-Antoine}, {Masana}, {Messineo}, {Mowlavi},
  {Nicolas}, {Nienartowicz}, {Pailler}, {Panuzzo}, {Riclet}, {Roux},
  {Seabroke}, {Sordo}, {Tanga}, {Th{\'e}venin}, {Gracia-Abril}, {Portell},
  {Teyssier}, {Altmann}, {Andrae}, {Bellas-Velidis}, {Benson}, {Berthier},
  {Blomme}, {Brugaletta}, {Burgess}, {Busso}, {Carry}, {Cellino}, {Cheek},
  {Damerdji}, {Davidson}, {Delchambre}, {Dell'Oro},
  {Fern{\'a}ndez-Hern{\'a}ndez}, {Galluccio}, {Garc{\'\i}a-Lario},
  {Garcia-Reinaldos}, {Gonz{\'a}lez-N{\'u}{\~n}ez}, {Gosset}, {Haigron},
  {Halbwachs}, {Hambly}, {Harrison}, {Hatzidimitriou}, {Heiter},
  {Hern{\'a}ndez}, {Hestroffer}, {Hodgkin}, {Holl}, {Jan{\ss}en}, {Jevardat de
  Fombelle}, {Jordan}, {Krone-Martins}, {Lanzafame}, {L{\"o}ffler}, {Lorca},
  {Manteiga}, {Marchal}, {Marrese}, {Moitinho}, {Mora}, {Muinonen}, {Osborne},
  {Pancino}, {Pauwels}, {Recio-Blanco}, {Richards}, {Riello}, {Rimoldini},
  {Robin}, {Roegiers}, {Rybizki}, {Siopis}, {Smith}, {Sozzetti}, {Ulla},
  {Utrilla}, {van Leeuwen}, {van Reeven}, {Abbas}, {Abreu}, {Accart}, {Aerts},
  {Aguado}, {Ajaj}, {Altavilla}, {{\'A}lvarez}, {{\'A}lvarez Cid-Fuentes},
  {Alves}, {Anderson}, {Anglada Varela}, {Antoja}, {Audard}, {Baines}, {Baker},
  {Balaguer-N{\'u}{\~n}ez}, {Balog}, {Barache}, {Barbato}, {Barros}, {Barstow},
  {Bartolom{\'e}}, {Bassilana}, {Bauchet}, {Baudesson-Stella}, {Becciani},
  {Bellazzini}, {Bernet}, {Bertone}, {Bianchi}, {Blanco-Cuaresma}, {Boch},
  {Bombrun}, {Bossini}, {Bouquillon}, {Bragaglia}, {Bramante}, {Breedt},
  {Bressan}, {Brouillet}, {Bucciarelli}, {Burlacu}, {Busonero}, {Butkevich},
  {Buzzi}, {Caffau}, {Cancelliere}, {C{\'a}novas}, {Cantat-Gaudin}, {Carballo},
  {Carlucci}, {Carnerero}, {Carrasco}, {Casamiquela}, {Castellani}, {Castro
  Sampol}, {Chaoul}, {Charlot}, {Chiavassa}, {Comoretto}, {Cooper}, {Cornez},
  {Cowell}, {Crifo}, {Crosta}, {Crowley}, {Dafonte}, {Dapergolas}, {David},
  {David}, {de Laverny}, {De Luise}, {De March}, {De Ridder}, {de Souza}, {de
  Teodoro}, {de Torres}, {del Peloso}, {del Pozo}, {Delgado}, {Delisle}, {Di
  Matteo}, {Diakite}, {Diener}, {Distefano}, {Dolding}, {Eappachen}, {Enke},
  {Esquej}, {Fabre}, {Fabrizio}, {Faigler}, {Fedorets}, {Fernique}, {Fienga},
  {Figueras}, {Fouron}, {Fragkoudi}, {Fraile}, {Franke}, {Gai}, {Garabato},
  {Garcia-Gutierrez}, {Garc{\'\i}a-Torres}, {Gavras}, {Gerlach}, {Geyer},
  {Giacobbe}, {Gilmore}, {Girona}, {Giuffrida}, {Gomez}, {Gonzalez-Santamaria},
  {Gonz{\'a}lez-Vidal}, {Granvik}, {Guti{\'e}rrez-S{\'a}nchez}, {Guy},
  {Hauser}, {Haywood}, {Hidalgo}, {Hilger}, {H{\l}adczuk}, {Hobbs}, {Holland},
  {Huckle}, {Jasniewicz}, {Jonker}, {Juaristi}, {Julbe}, {Karbevska},
  {Kervella}, {Khanna}, {Kochoska}, {Kontizas}, {Kordopatis}, {Korn},
  {Kostrzewa-Rutkowska}, {Kruszy{\'n}ska}, {Lambert}, {Lanza}, {Lasne}, {Le
  Campion}, {Le Fustec}, {Lebreton}, {Lebzelter}, {Leccia}, {Leclerc},
  {Lecoeur-Taibi}, {Liao}, {Licata}, {Lindstr{\o}m}, {Lister}, {Livanou},
  {Lobel}, {Madrero}, {Managau}, {Mann}, {Marchant}, {Marconi}, {Marcos},
  {Marinoni}, {Marocco}, {Marshall}, {Polo}, {Mart{\'\i}n-Fleitas}, {Masip},
  {Massari}, {Mastrobuono-Battisti}, {Mazeh}, {Messina}, {Michalik}, {Millar},
  {Mints}, {Molina}, {Molinaro}, {Moln{\'a}r}, {Montegriffo}, {Morbidelli},
  {Morel}, {Morris}, {Mulone}, {Munoz}, {Murphy}, {Musella}, {Noval},
  {Ord{\'e}novic}, {Orr{\`u}}, {Osinde}, {Pagani}, {Pagano}, {Palaversa},
  {Palicio}, {Panahi}, {Pawlak}, {Pe{\~n}alosa}, {Penttil{\"a}}, {Piersimoni},
  {Pineau}, {Plachy}, {Plum}, {Poggio}, {Poretti}, {Poujoulet}, {Pr{\v{s}}a},
  {Pulone}, {Racero}, {Ragaini}, {Rainer}, {Raiteri}, {Rambaux}, {Ramos},
  {Ramos-Lerate}, {Re}, {Regibo}, {Reyl{\'e}}, {Riva}, {Rixon}, {Robichon},
  {Robin}, {Roelens}, {Rohrbasser}, {Rowell}, {Royer}, {Rybicki}, {Sadowski},
  {Sagrist{\`a}}, {Sahlmann}, {Salgado}, {Salguero}, {Samaras}, {Sanchez},
  {Sanna}, {Santove{\~n}a}, {Sarasso}, {Schultheis}, {Sciacca}, {Segol},
  {Segovia}, {S{\'e}gransan}, {Semeux}, {Siddiqui}, {Siebert}, {Siltala},
  {Slezak}, {Smart}, {Solano}, {Solitro}, {Souami}, {Souchay}, {Spagna},
  {Spoto}, {Steele}, {Steidelm{\"u}ller}, {Stephenson}, {S{\"u}veges},
  {Szabados}, {Szegedi-Elek}, {Taris}, {Tauran}, {Taylor}, {Teixeira},
  {Thuillot}, {Tonello}, {Torra}, {Torra}, {Turon}, {Unger}, {Vaillant}, {van},
  {Vanel}, {Vecchiato}, {Viala}, {Vicente}, {Voutsinas}, {Weiler}, {Wevers},
  {Wyrzykowski}, {Yoldas}, {Yvard}, {Zhao}, {Zorec}, {Zucker}, {Zurbach}, \&
  {Zwitter}}]{gaiaetal2020b}
{Gaia Collaboration}, {Luri}, X., {Chemin}, L., {et~al.} 2020, arXiv e-prints,
  arXiv:2012.01771

\bibitem[{{Gaia Collaboration} {et~al.}(2016){Gaia Collaboration}, {Prusti},
  {de Bruijne}, {Brown}, {Vallenari}, {Babusiaux}, {Bailer-Jones}, {Bastian},
  {Biermann}, {Evans}, \& et~al.}]{gaiaetal2016}
{Gaia Collaboration}, {Prusti}, T., {de Bruijne}, J.~H.~J., {et~al.} 2016,
  \aap, 595, A1

\bibitem[{{Graczyk} {et~al.}(2020){Graczyk}, {Pietrzynski}, {Thompson},
  {Gieren}, {Zgirski}, {Villanova}, {Gorski}, {Wielgorski}, {Karczmarek},
  {Narloch}, {Pilecki}, {Taormina}, {Smolec}, {Suchomska}, {Gallenne},
  {Nardetto}, {Storm}, {Kudritzki}, {Kaluszynski}, \& {Pych}}]{graczyketal2020}
{Graczyk}, D., {Pietrzynski}, G., {Thompson}, I.~B., {et~al.} 2020, arXiv
  e-prints, arXiv:2010.08754

\bibitem[{{Kamann} {et~al.}(2018){Kamann}, {Bastian}, {Husser}, {Martocchia},
  {Usher}, {den Brok}, {Dreizler}, {Kelz}, {Krajnovi{\'c}}, {Richard},
  {Steinmetz}, \& {Weilbacher}}]{kamannetal2018}
{Kamann}, S., {Bastian}, N., {Husser}, T.~O., {et~al.} 2018, \mnras, 480, 1689

\bibitem[{{Niederhofer} {et~al.}(2018){Niederhofer}, {Cioni}, {Rubele},
  {Schmidt}, {Bekki}, {de Grijs}, {Emerson}, {Ivanov}, {Marconi}, {Oliveira},
  {Petr-Gotzens}, {Ripepi}, {van Loon}, \& {Zaggia}}]{niederhoferetal2018}
{Niederhofer}, F., {Cioni}, M.-R.~L., {Rubele}, S., {et~al.} 2018, \aap, 613,
  L8

\bibitem[{{Niederhofer} {et~al.}(2021){Niederhofer}, {Cioni}, {Rubele},
  {Schmidt}, {Diaz}, {Matijevic̆}, {Bekki}, {Bell}, {de Grijs}, {El
  Youssoufi}, {Ivanov}, {Oliveira}, {Ripepi}, {Subramanian}, {Sun}, \& {van
  Loon}}]{niederhoferetal2021}
{Niederhofer}, F., {Cioni}, M.-R.~L., {Rubele}, S., {et~al.} 2021, \mnras
  [\eprint[arXiv]{2101.09099}]

\bibitem[{{Parisi} {et~al.}(2015){Parisi}, {Geisler}, {Clari{\'a}},
  {Villanova}, {Marcionni}, {Sarajedini}, \& {Grocholski}}]{parisietal2015}
{Parisi}, M.~C., {Geisler}, D., {Clari{\'a}}, J.~J., {et~al.} 2015, \aj, 149,
  154

\bibitem[{{Parisi} {et~al.}(2009){Parisi}, {Grocholski}, {Geisler},
  {Sarajedini}, \& {Clari{\'a}}}]{parisietal2009}
{Parisi}, M.~C., {Grocholski}, A.~J., {Geisler}, D., {Sarajedini}, A., \&
  {Clari{\'a}}, J.~J. 2009, \aj, 138, 517

\bibitem[{{Piatti}(2017{\natexlab{a}})}]{p17c}
{Piatti}, A.~E. 2017{\natexlab{a}}, \apjl, 846, L10

\bibitem[{{Piatti}(2017{\natexlab{b}})}]{p17a}
{Piatti}, A.~E. 2017{\natexlab{b}}, \apjl, 834, L14

\bibitem[{{Piatti}(2017{\natexlab{c}})}]{p17b}
{Piatti}, A.~E. 2017{\natexlab{c}}, \mnras, 465, 2748

\bibitem[{{Piatti}(2018)}]{p2018}
{Piatti}, A.~E. 2018, \mnras, 477, 2164

\bibitem[{{Piatti}(2019)}]{piatti2019}
{Piatti}, A.~E. 2019, \apj, 882, 98

\bibitem[{{Piatti}(2021)}]{piatti2021}
{Piatti}, A.~E. 2021, \aap, 647, A11

\bibitem[{{Piatti} {et~al.}(2019){Piatti}, {Alfaro}, \&
  {Cantat-Gaudin}}]{piattietal2019}
{Piatti}, A.~E., {Alfaro}, E.~J., \& {Cantat-Gaudin}, T. 2019, \mnras, 484, L19

\bibitem[{{Piatti} \& {Bica}(2012)}]{pb12}
{Piatti}, A.~E. \& {Bica}, E. 2012, \mnras, 425, 3085

\bibitem[{{Piatti} {et~al.}(2018){Piatti}, {Cole}, \& {Emptage}}]{petal2018}
{Piatti}, A.~E., {Cole}, A.~A., \& {Emptage}, B. 2018, \mnras, 473, 105

\bibitem[{{Piatti} {et~al.}(2007){Piatti}, {Sarajedini}, {Geisler}, {Clark}, \&
  {Seguel}}]{petal07d}
{Piatti}, A.~E., {Sarajedini}, A., {Geisler}, D., {Clark}, D., \& {Seguel}, J.
  2007, \mnras, 377, 300

\bibitem[{{Pryor} \& {Meylan}(1993)}]{pm1993}
{Pryor}, C. \& {Meylan}, G. 1993, in Astronomical Society of the Pacific
  Conference Series, Vol.~50, Structure and Dynamics of Globular Clusters, ed.
  S.~G. {Djorgovski} \& G.~{Meylan}, 357

\bibitem[{{Ripepi} {et~al.}(2019){Ripepi}, {Molinaro}, {Musella}, {Marconi},
  {Leccia}, \& {Eyer}}]{ripepietal2018}
{Ripepi}, V., {Molinaro}, R., {Musella}, I., {et~al.} 2019, \aap, 625, A14

\bibitem[{{van der Marel} {et~al.}(2002){van der Marel}, {Alves}, {Hardy}, \&
  {Suntzeff}}]{vdmareletal2002}
{van der Marel}, R.~P., {Alves}, D.~R., {Hardy}, E., \& {Suntzeff}, N.~B. 2002,
  \aj, 124, 2639

\bibitem[{{Walker} {et~al.}(2006){Walker}, {Mateo}, {Olszewski}, {Bernstein},
  {Wang}, \& {Woodroofe}}]{walker2006}
{Walker}, M.~G., {Mateo}, M., {Olszewski}, E.~W., {et~al.} 2006, \aj, 131, 2114

\bibitem[{{Zivick} {et~al.}(2020){Zivick}, {Kallivayalil}, \& {van der
  Marel}}]{zivicketal2020}
{Zivick}, P., {Kallivayalil}, N., \& {van der Marel}, R.~P. 2020, arXiv
  e-prints, arXiv:2011.02525

\bibitem[{{Zivick} {et~al.}(2018){Zivick}, {Kallivayalil}, {van der Marel},
  {Besla}, {Linden}, {Koz{\l}owski}, {Fritz}, {Kochanek}, {Anderson}, {Sohn},
  {Geha}, \& {Alcock}}]{zivicketal2018}
{Zivick}, P., {Kallivayalil}, N., {van der Marel}, R.~P., {et~al.} 2018, \apj,
  864, 55

\end{thebibliography}


\end{document}